\documentclass[aps,pre,reprint,superscriptaddress,showpacs,preprintnumbers,amsmath,amssymb]{revtex4-1}
\usepackage{graphicx}
\graphicspath{{fig/}}
\usepackage{dcolumn}
\usepackage{bm}
\usepackage{color}

\begin{document}
\preprint{Draft}
\title{Scaling relations between viscosity and diffusivity in shear-thickening suspensions}
\author{Abhinendra Singh}
\affiliation{Department of Macromolecular Science and Engineering, Case Western Reserve University, Cleveland, Ohio 44106, USA}
\email[]{abhinendra.singh@case.edu}
\author{Kuniyasu Saitoh}
\affiliation{Department of Physics, Faculty of Science, Kyoto Sangyo University, Kyoto 603-8555, Japan}
\email[]{k.saitoh@cc.kyoto-su.ac.jp}
%
\date{\today}
\begin{abstract}
Dense suspensions often exhibit a dramatic response to large external deformation. The recent body of work has related this behavior to transition from an unconstrained lubricated to a constrained frictional state. Here, we use numerical simulations to study the flow behavior and shear-induced diffusion of frictional non-Brownian spheres in two dimensions under simple shear flow. We first show that \emph{both} viscosity $\eta$ and diffusivity $D/\dot{\gamma}$ of the particles increase at characteristic shear stress, which is associated with lubrication to frictional transition. Subsequently, we propose a one-to-one relation between viscosity and diffusivity using the length scale $\xi$ associated with the size of collective motions (rigid clusters) of the particles. We demonstrate that $\eta$ and $D/\dot{\gamma}$ are controlled by $\xi$ in two distinct flow regimes,\ i.e.\ in the frictionless and frictional states,
where the one-to-one relation is described as a crossover from $D/\dot{\gamma}\sim\eta$ ({frictionless}) to $\eta^{1/3}$ ({frictional}). 
We also confirm the proposed power laws are insensitive to the interparticle friction and system size.
\end{abstract}
\maketitle
%
\section{Introduction}
\label{sec:intro}
Suspension of rigid particles dispersed in a liquid 
are ubiquitous in natural~\cite{Coussot_1997}, geophysical~\cite{Jerolmack_2019},  industrial~\cite{Bridgwater_1993}, and biological~\cite{Maheshwari_2019} phenomena.
These suspensions often display non-Newtonian rheology, including shear thinning, normal stress differences, shear-thickening, or even jamming~\cite{Morris_2020,Denn_2014,guazzelli_2018,Ness_2022, Jamali_2020}.
A typical feature observed in these systems is the eventual divergence of the viscosity $\eta$
and a fluid-solid transition beyond a critical volume fraction often called the jamming volume fraction $\phi_J$~\cite{review-rheol0}.
In systems with particles, the precise value of $\phi_J$ depends sensitively on numerous microscopic details as well as the history of sample~\cite{pp0,pp2,pp3,pp4,pp5}.
Significantly, the jamming point is affected by the constraints on inter-particle relative motion (sliding, rolling, and twisting)~\cite{Singh_2020,Santos_2020} (originating from roughness~\cite{Pradeep_2021, Singh_2022, Singh_2022a}, interfacial chemistry~\cite{james2018interparticle,James_2019}), particle-shape~\cite{Donev_2004, Salerno_2018}, and polydispersity~\cite{Hopkins_2013, Guy_2020, Srivastava_2021}.
A striking phenomenon commonly observed in diverse systems such as cornstarch suspensions~\cite{Fall_2008} and other ideal suspensions~\cite{Brown_2009, guazzelli_2018, Singh_2022}, wherein the shear stress increases faster than linearly in response to the applied deformation rate is termed as Shear-thickening (ST).
At volume fractions $\phi$ far below the frictional jamming volume fraction $\phi_J^\mu$, the relative viscosity $\eta_r$ increases continuously with shear rate, and it termed as Continuous shear-thickening (CST); here, the relative viscosity $\eta_r (\phi,\dot{\gamma})/\eta_0$ is normalized by the solvent viscosity.
At volume fractions close enough to the jamming volume fraction  $\phi_C<\phi<\phi_J^\mu$, $\eta_r$ often increases abruptly by order of magnitude at a critical shear rate $\dot{\gamma}_c$. This phenomenon is termed as Discontinuous shear-thickening (DST)~\cite{Morris_2020}.
At volume fractions $\phi \ge \phi_J^{\mu}$, the suspension flows smoothly at low stresses but can be driven into a reversible solid-like jammed state. This phenomenon is called shear jamming (SJ)~\cite{Peters_2016, Singh_2019, Seto_2019, Singh_2020}.
The recent body of work~\cite{Guy_2018, Singh_2020, Pradeep_2021, Singh_2022, Singh_2022a} has linked both CST and DST to the activation of constraints that hinder relative particle motion as stress $\sigma$ is increased. 
These {\it{stress-activated}} constraints might arise from the Coulombic, static friction~\cite{Lin_2015, Guy_2015, Clavaud_2017, Hsu_2018, Seto_2013a, Mari_2014, Ness_2016, Kawasaki_2018, Singh_2020, Singh_2018, Mari_2019} or from combined effect of hydrodynamics and asperities~\cite{Jamali_2019}. At low-applied stresses below $\sigma^*$, steric repulsion keeps particle surfaces apart, maintaining the lubrication layer between particles and offering no constraints in the tangential pairwise motion. Above $\sigma^*$, the stabilization is overcome, leading to the formation of direct, frictional contacts~\cite{Seto_2013a, Mari_2014, mari_discontinuous_2015, Ness_2016, Clavaud_2017, Singh_2020}.
The mean-field models~\cite{Wyart_2014,Singh_2018, Guy_2018, Singh_2019, Singh_2020} capture this transition between lubricated, frictionless ``unconstrained" state to frictional ``constrained" state predict CST at low volume fractions $\phi$, DST at volume fractions $\phi_C<\phi<\phi_J^\mu$, and shear jamming (SJ) at $\phi \ge \phi_J^\mu$.

Stress-activated constraints on relative tangential motion between particles due to static friction lead to stabilization of load bearing force network under external deformation~\cite{Mari_2014, thomas2018microscopic, Boromand_2018, Singh_2020, thomas2020investigating, Sedes_2020, Gameiro_2020, Xu_2020, Edens_2021, Nabizadeh_2022}. 
%
%
With an increase in $\sigma$, the number of frictional contacts increases, leading to particle motion deviating from a pure affine trajectory to pass each other under applied shear.
This implies that the particle trajectories are more correlated at higher stresses suggesting an enhanced collective motion of particle groups, as reported previously~\cite{Ness_2016, Singh_2020}.
In the case of frictionless dry granular particles, dissipation due to non--affine velocity fluctuations have been shown to be connected to the emergence of the jamming behavior~\cite{Andreotti_2012}.
As jamming approaches from below, the rearrangement of marginally rigid, isostatic clusters, which eventually give way to long-range elastic interactions, dominates the flow.
The non-Brownian suspensions under shear present a particular case where these ideas can be applied to a continuously evolving high-viscosity fluid whose vicinity to the isostatic condition can be tuned by two independent variables: $\phi$ and $\sigma$.
Here we connect the concepts developed in dry granular systems to the rheology of sheared dense non-Brownian frictional suspensions to derive the relationship between rheology and microscopic variables, namely diffusion of particles and non-affine velocity correlation length or so-called {\it rigid clusters}.

In this paper, we numerically study the rheology and diffusion of frictional non-Brownian spheres under simple shear.
We investigate the microstructure and dynamics as the suspension shear thickens; our specific aim is to extract the relations among viscosity, diffusivity, and correlation length of the particle motion.
In the following, we explain our numerical methods (Sec.\ \ref{sec:method}),
show our results of viscosity, diffusivity, and spatial correlations of the particles (Sec.\ \ref{sec:result}),
and discuss our findings and future work (Sec.\ \ref{sec:disc}).
%
\section{Numerical methods}
\label{sec:method}
We perform quasi-2D simple shear simulations with a monolayer of $N=2000$ non-Brownian spheres immersed in a Newtonian fluid.
To avoid crystallization of the system, we randomly distribute an equal volume of small and large particles (with radii $a_S=a$ and $a_L=1.4a$) in an $L\times L$ periodic square box using Lees-Edwards periodic boundary condition~\cite{lees,gn1}.
In the limit of zero Reynolds and Stokes numbers, the problem reduces to the overdamped limit and force balance between hydrodynamic $\mathrm{\mathbf{F}_H}$ and contact $\mathrm{\mathbf{F}_C}$ forces on every particle
\begin{equation}
    0 = \mathrm{\mathbf{F}_H}(\mathbf{r,u})+\mathrm{\mathbf{F}_C}(\mathbf{r}),
\end{equation}
where $\mathbf{r}$ and $\mathbf{u}$ are many body particle positions and velocities, respectively with $\mathbf{u}  \equiv \dot{\mathbf{r}}$.
Similar relation also holds for the torque balance.
We regularize the lubrication singularity at a surface separation $h=10^{-3}a$ to allow the contact between particles~\cite{Mari_2014}.
The contact between particles is modeled using the approach of Cundall and Strack~\cite{Cundall_1979} by introducing normal and tangential springs with spring constants being $k_n$ and $k_t=0.5k_n$, respectively.
We tune stiffness at every $\phi$ such that the maximum overlap does not exceed 3\% of the particle radius to stay near the rigid limit~\cite{Singh_2015}. 
To introduce rate dependence, a critical load model (CLM) scheme, where a critical normal force $F^*$ is needed to activate friction between particles~\cite{Mari_2014}.
Here, we use interparticle friction coefficient $\mu=1$.
Note that we examine the effects of the system size $N$ and friction coefficient $\mu$ on our numerical results in Appendix \ref{app:depend}.

To simulate simple shear flows of the system, we apply external shear stress $\sigma$ to the system under the Lees-Edwards boundary condition \cite{lees}.
The inset to Fig.\ \ref{fig:DST_snapshot_msdy_corl}a displays a snapshot of our simulation, where the system is sheared along the horizontal arrows.
At constant shear stress, the shear rate becomes the fluctuating time-dependent quantity,
\begin{equation}
\dot{\gamma} = \frac{\sigma-\sigma_{\mathrm C}}{\eta_0(1+2.5\phi)+\eta_{\mathrm H}}~,
\end{equation}
where $\sigma_{\mathrm C}$ and $\sigma_{\mathrm H}$ are the contact and hydrodynamic contributions to the total stress~\cite{Mari_2015,Seto_2019}.
All the length scales are normalized by small particle radius $a$.
The unit scale for shear rate and stress are 
$\dot{\gamma}_0 \equiv F^\ast/ 6\pi \eta_0 a^2$ and $\sigma_0 \equiv \eta_0 \dot{\gamma}_0$, respectively.
The time scale in the simulations is $t_0 = 6 \pi a \eta_0 / F^\ast$.
Here $F^\ast$ is the minimum force needed to activate frictional contacts leading to shear-thickening behavior.

From the simulations, the particle positions, normal and tangential contact forces, and non-contact lubrication forces are extracted.
At any given condition, the simulations are performed for $40$ strain units, and the data are recorded in strain steps of $10^{-1}$.
%
\section{Results}
\label{sec:result}
This section shows our numerical results of the quasi-2D simple shear simulations.
First, we focus on transverse motions of the non-Brownian particles under shear,
where their diffusivity and correlation length are extracted from the transverse component of \emph{mean squared displacement} (MSD)
and spatial correlation function in transverse velocities, respectively (Sec.\ \ref{sub:trans}).
Further, we examine how the diffusivity and correlation length are affected by increase in volume fraction and stress (Sec.\ \ref{sub:rheol}).
Then, we relate the viscosity to diffusivity by scaling arguments (Sec.\ \ref{sub:scaling}),
where the scaling relations are robust to the changes of friction coefficient $\mu$ and system size $N$ (Appendix B).
\subsection{Transverse motions}
\label{sub:trans}
\begin{figure}
\includegraphics[width=\columnwidth]{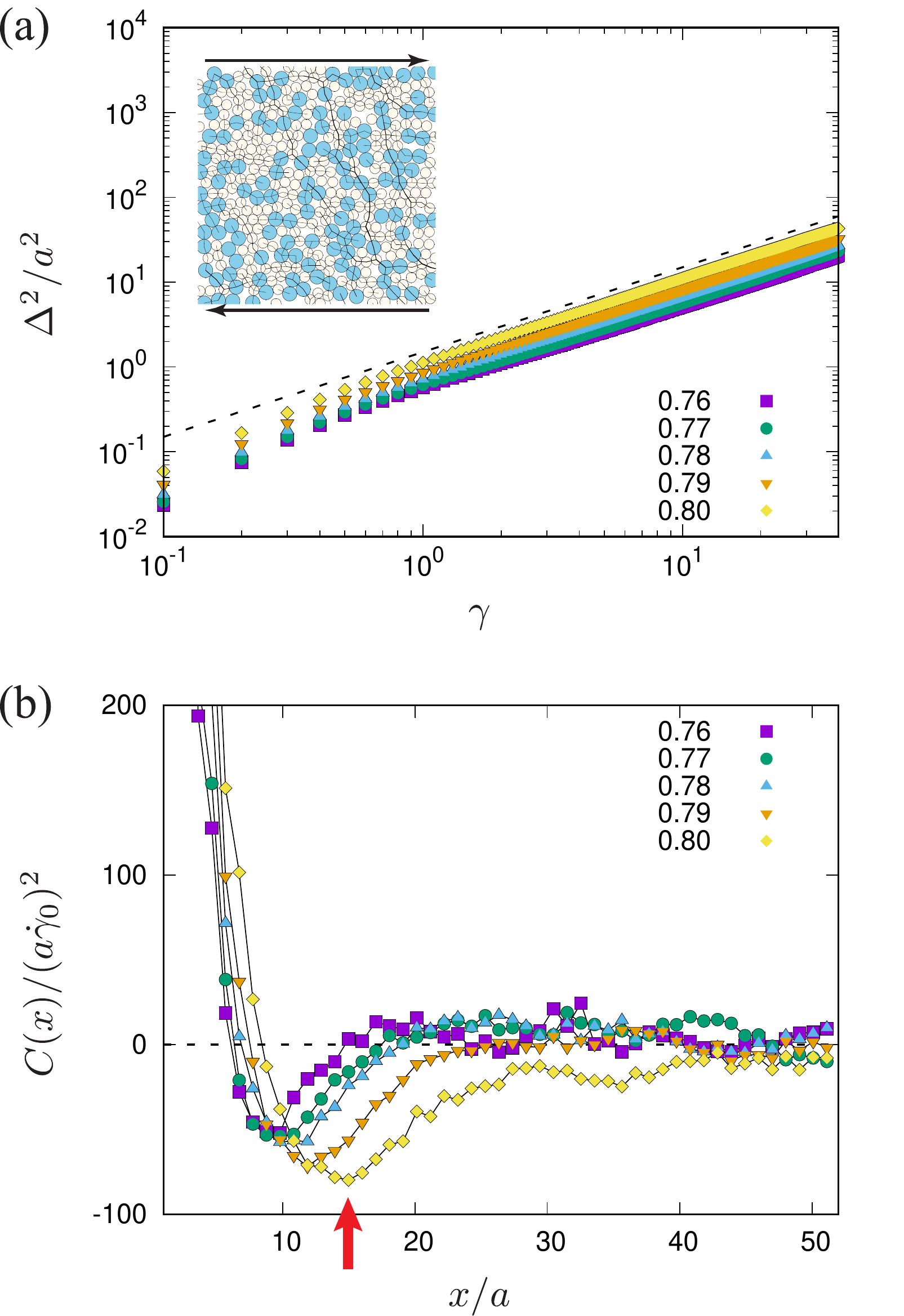}
\caption{
(a) Transverse MSDs, $\Delta(\tau)^2$ (Eq.\ (\ref{eq:MSD})), as functions of the shear strain $\gamma\equiv\dot{\gamma}\tau$,
where the dashed line indicates the normal diffusion, $\Delta^2\sim\gamma$.
The packing fraction $\phi$ increases as listed in the legend.
Inset: A snapshot of our MD simulation, where the particles (circles) are sheared along the horizontal arrows, and the lines represent force-chain networks.
(b) Spatial correlation functions of the transverse velocities $C(x)$ (Eq.\ (\ref{eq:C(x)})).
The vertical arrow indicates the minimum of $C(x)$ for the data of $\phi=0.80$.
The symbols are as in (a).
\label{fig:DST_snapshot_msdy_corl}}
\end{figure}

We analyze diffusion of the particles under shear by the transverse component of MSD \cite{diff_shear_md0,diff_shear_md1,diff_shear_md3,diff_shear_md4},
\begin{equation}
\Delta(\tau)^2 = \left\langle\frac{1}{N}\sum_{i=1}^N\Delta y_i(\tau)^2\right\rangle~.
\label{eq:MSD}
\end{equation}
Here, $\Delta y_i(\tau)$ is the $y$-component of particle displacement for the duration $\tau$
and the ensemble average $\langle\dots\rangle$ is taken over different choices of the initial time
\footnote{The MSDs defined by the \emph{total} non-affine displacements show quantitatively the same results (data are not shown for brevity).}.
Figure \ref{fig:DST_snapshot_msdy_corl}a displays the typical transverse MSDs, $\Delta(\tau)^2$,
where the horizontal axis is the duration scaled by shear rate, $\gamma\equiv\dot{\gamma}\tau$\, i.e., \ the shear strain applied to the system for the duration $\tau$.
As can be seen, the MSD exhibits a crossover from super-diffusion ({in the sense that the slope is larger than unity}) to normal diffusion (dashed line) around $\gamma\approx 1$.
This trend is qualitatively insensitive to the packing fraction $\phi$ and stress $\sigma$ (though the MSD increases with the increase of $\phi$, as listed in the legend).
Note that, different from thermal systems under shear \cite{rheol10,nafsc5,th-dh_md1}, we do not observe any plateaus in the MSDs.
Therefore, neither \emph{caging} nor \emph{sub-diffusion} of the particles exists in our system \cite{dh_md2,dh_qs1,dh_md1}.

We further introduce a correlation length as a typical size of \emph{rigid clusters} \cite{gn5,rheol0,nafsc2,diff_shear_md1,saitoh11,saitoh14},
where collective motions of the particles are interpreted as a result of rigid body rotations of the clusters.
Following Ref.~\cite{rheol0,pdf1,corl3}, we quantify the collective motions by a spatial correlation function,
\begin{equation}
C(x)=\langle v_y(x_i,y_i)v_y(x_i+x,y_i)\rangle~,
\label{eq:C(x)}
\end{equation}
where $v_y(x,y)$ is the transverse velocity field, and the ensemble average $\langle\dots\rangle$ is taken over particle positions and time (in the steady state).
Figure \ref{fig:DST_snapshot_msdy_corl}b shows the correlation functions, where the horizontal axis ($x$-axis) is scaled by the small particle radius $a$.
As can be seen, the correlation function exhibits a well-defined minimum at a characteristic distance, $x=\xi$ (as indicated by the vertical arrow for the data of $\phi=0.80$).
Because the minimum is negative, $C(\xi)<0$, the transverse velocities are most ``anti-correlated" at the minima $x=\xi$ \cite{rheol0}.
Assuming that the rigid cluster is circular, its mean diameter is comparable in size with $\xi$
\footnote{
In a short range of $x$, transverse velocities tend to be aligned in the same direction,\ i.e., \ $v_y(x_i,y_i)v_y(x_i+x,y_i)>0$,
such that the correlation function $C(x)$ is a positive decreasing function of $x$.
If the transverse velocities align in the opposite direction,\ i.e.\ $v_y(x_i,y_i)v_y(x_i+x,y_i)<0$, the correlation function becomes negative, $C(x)<0$.
Because the correlation function is minimum when the transverse velocities are located on either side of a vortex-like structure, typical size of rigid clusters can be defined as the distance at which $C(x)$ becomes minimum.
In a long distance of $x$, due to the randomness of transverse velocities, the correlation function eventually decays to zero \cite{saitoh11}.}.
Note that this ``rigid cluster" is different from the rigid clusters extracted using various algorithms like pebble game on the contact network~{\cite{Silke_2016, Zhang_2019}}.
%
\subsection{Viscosity, diffusivity, and size of rigid clusters}
\label{sub:rheol}
\begin{figure*}
\includegraphics[width=\textwidth]{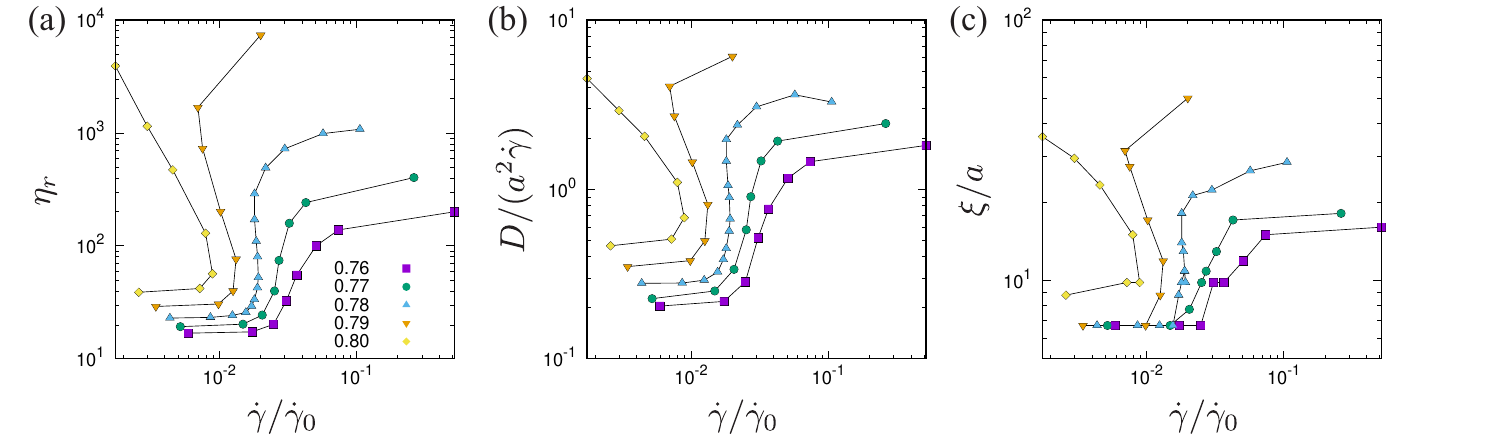}
\caption{
Double logarithmic plots of (a) the relative viscosity $\eta_r$, (b) diffusivity (Eq.\ (\ref{eq:D})) over the shear rate $D/\dot{\gamma}$,
and (c) size of rigid clusters $\xi$ as functions of the shear rate $\dot{\gamma}$.
Different symbols represent different values of the packing fraction $\phi$ as listed in the legend of (a).
\label{fig:DST_et_df_xi}}
\end{figure*}

%
Figure \ref{fig:DST_et_df_xi}a displays $\eta_r$ as functions of the shear rate $\dot{\gamma}$ for packing fraction $\phi$ as listed in the legend.
For $\phi \le 0.77$, the shear-thickening is continuous between two rate-independent values,\, i.e., CST is observed.
The shear thickening becomes discontinuous for $\phi\rightarrow\phi_c\simeq 0.78$ and $\eta_r$ exhibits the characteristic ``S-shape" for $\phi \ge 0.78$.
For $\phi=0.80$, the system at the highest stress, $\sigma/\sigma_0=100$, is eventually \emph{shear jammed} such that the viscosity cannot be defined.
Note that the simulations presented here are performed under fixed shear stress $\sigma$,
while the S-shape indicates a discontinuous jump (or drop) of $\eta_r$ in rate-controlled experiments \cite{Boersma_1991}, and simulations \cite{Mari_2015}.
%

As shown in Fig.\ \ref{fig:DST_snapshot_msdy_corl}a, the diffusion of the particles becomes normal diffusion, where $\Delta^2\sim\gamma$ (dashed line) for $\gamma>1$.
To quantify the normal diffusion of the particles, we introduce diffusivity as
\footnote{We define the diffusivity [Eq.\ (\ref{eq:D})] as the slope of the MSD [Eq.\ (\ref{eq:MSD})] in the normal diffusive regime $\gamma=\dot{\gamma}\tau>1$,
where we take sample averages of $\Delta(\tau)^2/2\tau$ as $D/\dot{\gamma}\equiv <\Delta(\gamma)^2/2\gamma>$ in the range between $1<\gamma<40$.}
\begin{equation}
D=\lim_{\tau\rightarrow\infty}\frac{\Delta(\tau)^2}{2\tau}~.
\label{eq:D}
\end{equation}
Figure \ref{fig:DST_et_df_xi}b shows diffusivity scaled by the shear rate, $D/\dot{\gamma}$, as functions of the shear rate $\dot{\gamma}$.
The diffusivity of particles exhibits behavior across the thickening transition analogous to rheology.
We observe a rate-independent plateau in the frictionless and frictional limits, with increased diffusivity as the suspension shear thickens.
Note that diffusivity $D/\dot{\gamma}$ cannot be defined at the highest stress ($\sigma/\sigma_0=100$) if $\phi=0.80$,
where the system is shear jammed so that no particle can diffuse (in the transverse $y$-direction).

To further probe particle scale dynamics, we analyze the typical size of rigid clusters $\xi$ using the correlation function, {$C(x)$}.
Figure \ref{fig:DST_et_df_xi}c plots $\xi$ as functions of scaled shear rate $\dot{\gamma}/\dot{\gamma}_0$ for various values of $\phi$.
Again, its dependence on $\phi$ and $\dot{\gamma}$ is quite similar to those of $\eta_r$ and $D/\dot{\gamma}$ (Figs.\ \ref{fig:DST_et_df_xi}a and b):
$\xi$ increases as the suspension shear thickens.
For $\phi=0.80$ the system is shear jammed (and thus does not induce collective motions of the particles) \cite{diff_shear_md1,diff_shear_md8} at the highest stress ($\sigma/\sigma_0=100$).

In Appendix A, we analyze the \emph{coordination number} $z$ and \emph{anisotropic fabric} $A$ of the particles \cite{Ness_2016}.
The coordination number $z$ increase with increasing stress. At the same time, the contact network becomes less anisotropic as the suspension shear thickens (see Appendix A).
These observations are consistent with the findings of Ness \& Sun~\cite{Ness_2016}.
%

\subsection{Scaling {arguments}}
\label{sub:scaling}
In the previous section, we have shown that the dependence of diffusivity of particles $D/\dot{\gamma}$ on $\phi$ and $\dot{\gamma}/\dot{\gamma}_0$ is similar to that of $\eta_r(\phi,\dot{\gamma}/\dot{\gamma}_0)$.
Motivated by this, below we derive relation between the rheology and diffusion of the non-Brownian particles,
which is reminiscent of the Stokes-Einstein relation for Brownian particles.
To reveal such a relation, we connect the viscosity $\eta_r$ with diffusivity $D$ based on the scaling arguments in Refs.\ \cite{CST0,CST1,diff_shear_md1,diff_shear_md8,saitoh14}.

\begin{figure}
\includegraphics[width=\columnwidth]{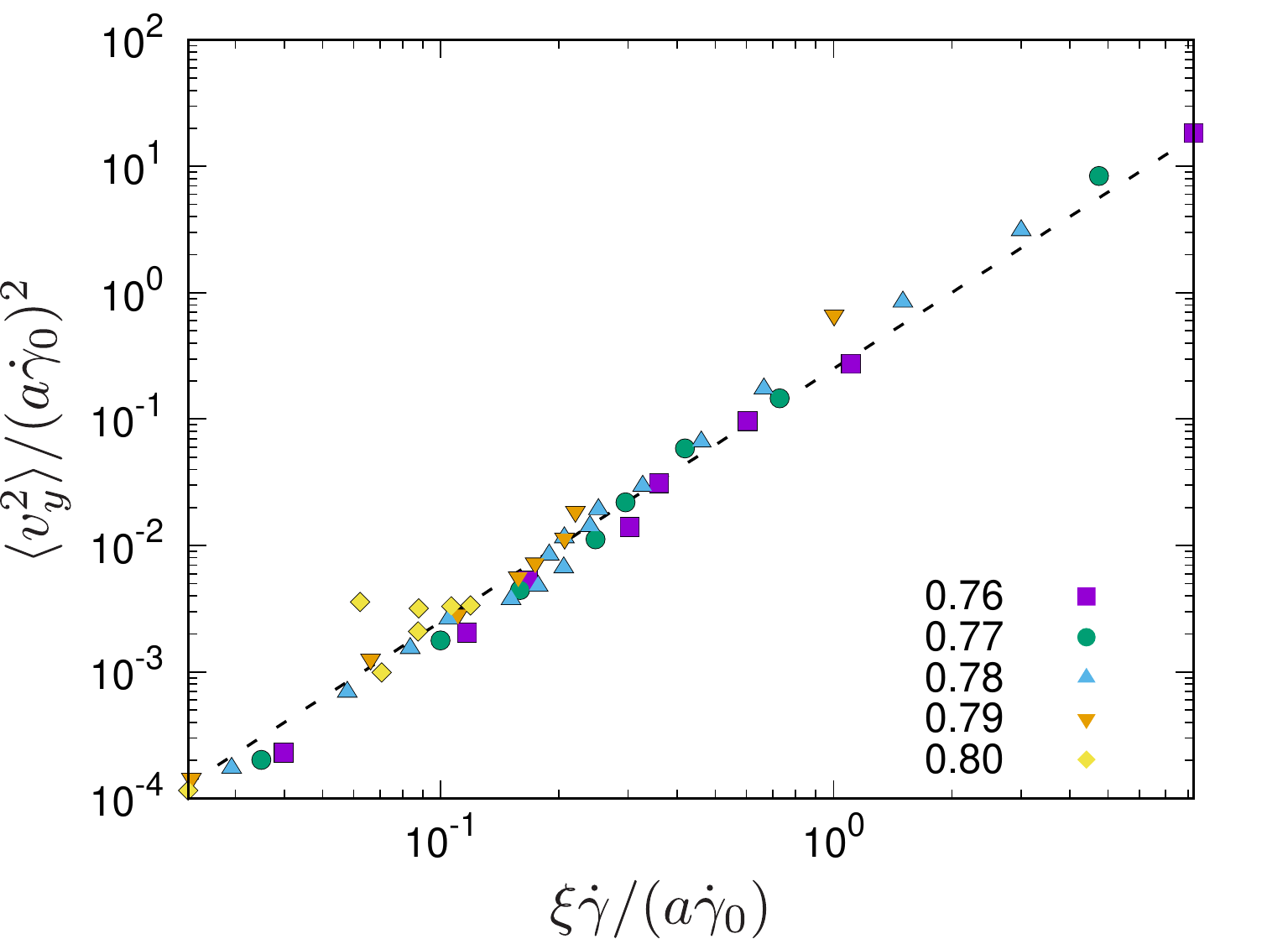}
\caption{
Scatter plots of $\langle v_y^2\rangle$ and $\xi\dot{\gamma}$,
where the dashed line represents the scaling, $\langle v_y^2\rangle\sim (\xi\dot{\gamma})^2$.
The packing fraction $\phi$ increases as listed in the legend.
\label{fig:DST_fluctuations_length}}
\end{figure}

Let us represent the transverse speed of the particles by characteristic length and time scales as \cite{CST0,CST1}
\begin{equation}
v_y \sim \xi\dot{\gamma}~.
\label{eq:delta_v}
\end{equation}
To examine Eq.\ (\ref{eq:delta_v}), we plot $\langle v_y^2\rangle$ as a function of $\xi\dot{\gamma}$, where
\begin{equation}
\langle v_y^2\rangle = \left\langle\frac{1}{N}\sum_{i=1}^N v_{iy}^2\right\rangle
\end{equation}
is the mean squared transverse speed ($\langle\dots\rangle$ is the time average in a steady state).
Figure\ \ref{fig:DST_fluctuations_length} confirms the scaling, Eq.\ (\ref{eq:delta_v}), over the whole ranges of $\phi$ and $\dot{\gamma}$,
where the dashed line indicates $\langle v_y^2\rangle \sim (\xi\dot{\gamma})^2$.

To extract the dependence of $\eta_r$ on the length scale $\xi$ we use the energy balance equation \cite{CST0,CST1}.
As the system reaches a steady state, the injection of energy by shear $\dot{\gamma}\sigma$
and the rate of energy dissipation $\Gamma$ (due to dissipative forces) needs to be in balance as $\dot{\gamma}\sigma\sim\Gamma$ \cite{energydissipation}.
If the shear stress (or equivalently shear rate) is low enough, the system is in the frictionless state
where the drag force is more significant for energy dissipation than the contact damping so that $\Gamma\approx\Gamma_\mathrm{visc}\sim v_y^2$
\footnote{
The dissipation rate due to the drag force is proportional to the square of velocity fluctuation, $\Gamma_\mathrm{visc}\propto (\delta v)^2$ \cite{CST1}.
Because the mean flow is parallel to the $x$-axis, the velocity fluctuation is comparable in size with the transverse speed, $\delta v\sim v_y$,
and thus $\Gamma\approx\Gamma_\mathrm{visc}\sim v_y^2$.}.
At higher stresses, as the frictional contacts form, interparticle friction becomes the dominant source of energy dissipation,
where $\Gamma\approx\Gamma_\mathrm{fric}\sim v_y^3/\dot{\gamma}$
\footnote{
The dissipation rate due to the friction scales as $\Gamma_\mathrm{fric}\propto \xi\delta v$ \cite{CST1}.}.
Substituting Eq.\ (\ref{eq:delta_v}) into $\Gamma_\mathrm{visc}$ and $\Gamma_\mathrm{fric}$,
we find scaling relations of the dissipation rate as \cite{CST1}
\begin{equation}
\Gamma \sim
\begin{cases}
\xi^2\dot{\gamma}^2 & (\mathrm{{frictionless}}) \\
\xi^3\dot{\gamma}^2 & (\mathrm{{frictional}})
\end{cases}~.
\label{eq:Gamma}
\end{equation}
If we rewrite the energy injection as $\dot{\gamma}\sigma=\dot{\gamma}^2\eta_r$,
the energy balance ($\dot{\gamma}\sigma \sim \Gamma$) tells us $\eta_r \sim\Gamma/\dot{\gamma}^2$ so that the viscosity depends on the length scale as
\begin{equation}
\eta_r \sim
\begin{cases}
\xi^2 & (\mathrm{{frictionless}}) \\
\xi^3 & (\mathrm{{frictional}})
\end{cases}~.
\label{eq:eta}
\end{equation}

In addition, it was recently pointed out that rotations of rigid clusters govern the diffusion of the particles under shear,
where the size of rigid clusters describes the diffusivity as $D\sim a\xi\dot{\gamma}$ \cite{diff_shear_md1,diff_shear_md8,saitoh14}.
At low stresses, the suspension is in the frictionless state; the relative particle motion is correlated to only a few particles radii (Fig.\ \ref{fig:DST_et_df_xi}c),\ i.e.\ $\xi\sim a$,
meaning that the diffusivity is simply given by $D\sim\xi^2\dot{\gamma}$.
As frictional contacts form with increasing stress, the correlation length can be as large as $\xi\sim 50a$ (Fig.\ \ref{fig:DST_et_df_xi}c).
This presents a length scale associated with particle dynamics that is distinct from the particle size,\ {i.e.\ $\xi\gg a$}.
Additionally diffusivity scales as $D\sim\xi\dot{\gamma}$ \cite{diff_shear_md1,diff_shear_md8,saitoh14}.
To summarize, scaling relations of the diffusivity are given by
\begin{equation}
D \sim
\begin{cases}
\xi^2\dot{\gamma} & (\mathrm{{frictionless}}) \\
\xi\dot{\gamma} & (\mathrm{{frictional}})
\end{cases}~.
\label{eq:scaling_D}
\end{equation}

\begin{figure}
\includegraphics[width=\columnwidth]{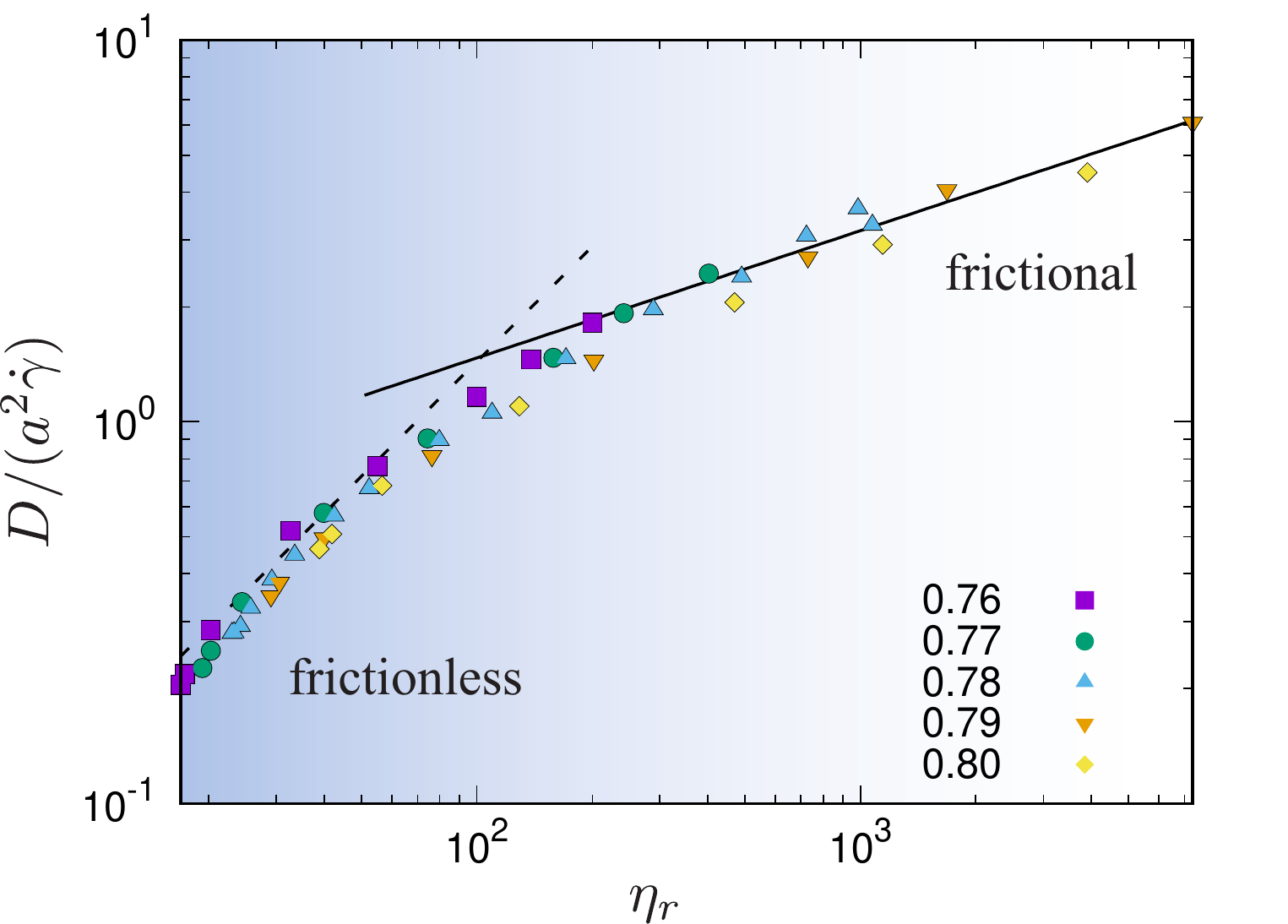}
\caption{
Scatter plots of the diffusivity over the shear rate $D/\dot{\gamma}$ and relative viscosity $\eta_r$.
The dashed line indicates the scaling in the {frictionless state},\ i.e.\ $D/\dot{\gamma}\sim\eta_r$,
while the solid line is that in the {frictional} state,\ i.e.\ $D/\dot{\gamma}\sim\eta_r^{1/3}$.
The packing fraction $\phi$ increases as listed in the legend.
\label{fig:DST_viscosity_diffusivity}}
\end{figure}

From Eqs.\ (\ref{eq:eta}) and (\ref{eq:scaling_D}), we can derive a connection between the viscosity and diffusivity as
\begin{equation}
D/\dot{\gamma} \sim
\begin{cases}
\eta_r & (\mathrm{{frictionless}}) \\
\eta_r^{1/3} & (\mathrm{{frictional}})
\end{cases}~.
\label{eq:D_gd_eta}
\end{equation}
To validate Eq.\ (\ref{eq:D_gd_eta}), we plot $D/\dot{\gamma}$ vs $\eta_r$ in Fig.\ \ref{fig:DST_viscosity_diffusivity},
where the dashed and solid lines indicate the relations, $D/\dot{\gamma} \sim \eta_r$ and $\eta_r^{1/3}$, respectively.

In Appendix B, we show that one-to-one relations,\ Eq.\ (\ref{eq:D_gd_eta}), are robust to the changes of friction coefficient $\mu$ and system size $N$.
Therefore, it is universal that the viscosity and diffusivity of the non-Brownian particles under shear are governed by the size of rigid clusters and are related with each other.
Because the one-to-one relations are insensitive to the system size,
our results are distinct from numerical results of sheared amorphous solids \cite{diff_shear_md3,diff_shear_md4},
where the diffusivity over the shear rate becomes linear in the system length as $D/\dot{\gamma}\sim L$ in the quasi-static limit.
In Refs.\ \cite{diff_shear_md3,diff_shear_md4}, the interparticle friction is not introduced
so that S-shaped transitions of $\eta_r$ and $D/\dot{\gamma}$ (from the frictionless to frictional states) are not expected.
Thus, the one-to-one relations,\ Eq.\ (\ref{eq:D_gd_eta}), are specific to our frictional non-Brownian particles.
%
\section{Discussion}
\label{sec:disc}
In this study, we have numerically investigated the rheology, diffusion, and spatial correlations of rigid frictional non-Brownian suspension under shear.
We analyzed the transverse component of MSD and spatial correlation function of the transverse velocities
to extract the diffusivity over the shear rate $D/\dot{\gamma}$ and size of rigid clusters $\xi$, respectively.
We found that the dependence of $D/\dot{\gamma}$ and $\xi$ on the packing fraction $\phi$ and shear rate $\dot{\gamma}$ are analogous to that of the relative viscosity $\eta_r(\dot{\gamma})$.
They tend to plateau at low and high shear rates (stresses) and increase between these plateaus as the suspension shear thickens.
We have derived one-to-one {relations} between rheology and diffusion using extensive numerical data for rheology, diffusivity, and correlation lengths.
This, in spirit, is similar to the Stokes-Einstein relation for Brownian particles.
Additionally, making use of the scaling arguments suggested in Refs.\ \cite{CST0,CST1},
we introduced the power-law scaling $\eta_r\sim\xi^\alpha$ with the exponents, $\alpha=2$ and $3$, for the {frictionless} and {frictional states}, respectively.
We also explained the scaling relations of the diffusivity as $D\sim\xi^2\dot{\gamma}$ ({frictionless}) and $\xi\dot{\gamma}$ ({frictional}).
Then, from these scaling relations, we have {found} the one-to-one relations as $D/\dot{\gamma}\sim\eta_r$ and $\eta_r^{1/3}$ for the {frictionless} and {frictional states}, respectively.
These relations were numerically validated and found to be robust to the changes of the friction coefficient $\mu$ and system size $N$.
{
It is surprising that the one-to-one relations,\ Eq.\ (\ref{eq:D_gd_eta}), are not altered by the changes of $\mu$ and $N$
because the shear-thickening is sensitive to the interparticle friction \cite{Seto_2013a} and the diffusivity is believed to be system-size dependent \cite{diff_shear_md3,diff_shear_md4}.
Our results suggest that the typical size of rigid clusters $\xi$ exhibits a similar dependence on the parameters ($\phi$, $\mu$, and $N$) with $\eta_r$ and $D/\dot{\gamma}$,
and the viscosity and diffusivity of frictional non-Brownian particles are governed by the length scale $\xi$.}
\subsection{Comparison with previous studies}
In Refs.\ \cite{CST0,CST1}, the scaling relations,\ Eqs.\ (\ref{eq:delta_v}) and (\ref{eq:Gamma}), have been numerically confirmed for the 2D system of inertial suspensions under simple shear.
We also numerically validated Eq.\ (\ref{eq:delta_v}) and used Eq.\ (\ref{eq:Gamma}) to derive the one-to-one relations,\ Eq.\ (\ref{eq:D_gd_eta}).
The main difference between our model and that in Refs.\ \cite{CST0,CST1} is the absence (or presence) of particle inertia.
Therefore, the scaling relations,\ Eqs.\ (\ref{eq:delta_v}) and (\ref{eq:Gamma}), seem to be independent of the particle mass
though it is left to future work to examine whether the one-to-one relations,\ Eq.\ (\ref{eq:D_gd_eta}), are altered by the particle inertia.

In Ref.\ \cite{Ness_2016}, the sudden increases of $D/\dot{\gamma}$, $\xi$, coordination number $z$, and anisotropic fabric $A$ around DST have already been reported.
However, the connection between rheology and diffusion was not investigated, and the effects of friction coefficient $\mu$ and system size $N$ were not studied.
In addition, it was found in molecular dynamics simulations of sheared amorphous solids \cite{diff_shear_md3,diff_shear_md4}
that the diffusivity over the shear rate becomes linear in the system length as $D/\dot{\gamma}\sim L$
if the packing fraction is far above the jamming transition density and the shear rate is sufficiently small.
Different from the numerical model used in Refs.\ \cite{diff_shear_md3,diff_shear_md4}, {we use} the frictional non-Brownian particles {and} the correlation length $\xi$ does not extend to the system length,\, i.e., \ $\xi<L$.
As a result, we did not observe any finite-size effects in {the diffusivity}.
\subsection{Future works}
In our simulations, we employed the model for shear-thickening suspensions to investigate the connection between their rheology and diffusion under shear.
However, another drastic change of the viscosity is known as \emph{discontinuous shear thinning} \cite{AST2},
where cohesive forces are introduced between the frictionless particles.
Therefore, it is important to analyze the relation between rheology and diffusion of soft cohesive particles under shear
and clarify whether the diffusion is ``suppressed" by discontinuous shear thinning or if cohesion would lead to shear localization~\cite{Singh_2014}.
In addition, recent studies of the rheology of non-spherical particles \cite{nsphere0,nsphere1,nsphere2}
suggest that the influence of particle shapes on one-to-one relations should be examined in the future.
Moreover, extending the current work to three dimensions \cite{rheol15,saitoh16} is crucial to the practical applications of this work.
Confinement that can slow down the dynamics~\cite{Li_2020, Singh_2020s} is another essential aspect to investigate.
\begin{acknowledgments}
We thank A. Ikeda, H. Mizuno, and S. Yamanaka for fruitful discussions.
A. S. would like to acknowledge Case Western Reserve University for the start-up funding for the project.  
K.S. was financially supported by JSPS KAKENHI from JSPS (Grant Numbers 18K13464, 20H01868, and 21H01006) and Inamori Research Grants 2021.
\end{acknowledgments}
%

\bibliography{SE_relation,dst}

\clearpage

\appendix
\section{Viscosity, diffusivity, and size of rigid clusters as functions of the external shear stress}
\label{app:contact}
In this appendix, we show additional data of the relative viscosity $\eta_r$, diffusivity over the shear rate $D/\dot{\gamma}$, and size of rigid clusters $\xi$.
Since we apply the external shear stress $\sigma$ to the system, the shear rate $\dot{\gamma}$ is ``measured" in simulations.
We plot $\eta_r$, $D/\dot{\gamma}$, and $\xi$ as functions of $\sigma$ which is the control parameter.
The evolution of \emph{contact networks} is presented using coordination number $z$ and anisotropic fabric $A$ \cite{Ness_2016}.

\begin{figure}
\includegraphics[width=\columnwidth]{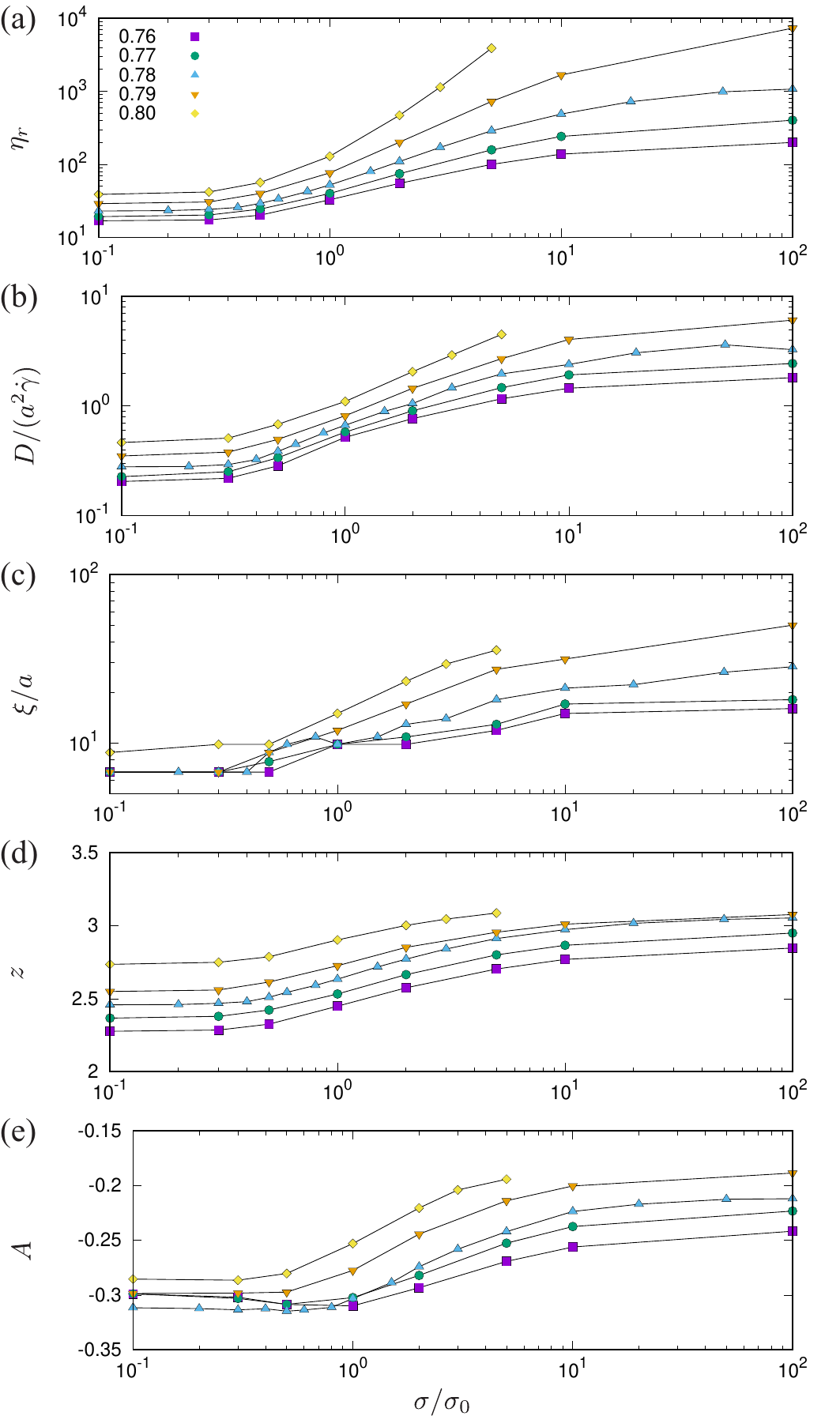}
\caption{
Double-logarithmic plots of (a) the relative viscosity $\eta_r$, (b) diffusivity over the shear rate $D/\dot{\gamma}$,
and (c) size of rigid clusters $\xi$ as functions of the external shear stress $\sigma$.
Semi-logarithmic plots of (d) the coordination number $z$ and (e) anisotropic fabric $A$ as functions of $\sigma$.
Here, the packing fraction $\phi$ increases as listed in the legend of (a).
\label{fig:DST_et_df_xi_sz_sb-se}}
\end{figure}

Figures \ref{fig:DST_et_df_xi_sz_sb-se}a-c are double-logarithmic plots of (a) $\eta_r$, (b) $D/\dot{\gamma}$, and (c) $\xi$ as functions of $\sigma$,
where the packing fraction $\phi$ increases as listed in the legend of (a).
Increasing $\sigma$, we observe that all the quantities start to increase around a characteristic shear stress, $\sigma=\sigma_0 = 0.3$.
In these figures, the data of the densest packing ($\phi=0.80$) are truncated at $\sigma=5\sigma_0$
because the system is shear-jammed,\ i.e.\ does not flow, for $\sigma>5\sigma_0$ so that $\eta_r$, $D/\dot{\gamma}$, and $\xi$ cannot be defined.
Figures \ref{fig:DST_et_df_xi_sz_sb-se}d and e are semi-logarithmic plots of (d) $z$ and (e) $A$ as functions of $\sigma$.
As can be seen, both $z$ and $A$ increase as the suspension shear-thickens,
implying that as frictional contacts form, the contact network rearranges and becomes less anisotropic.

\begin{figure}
\includegraphics[width=\columnwidth]{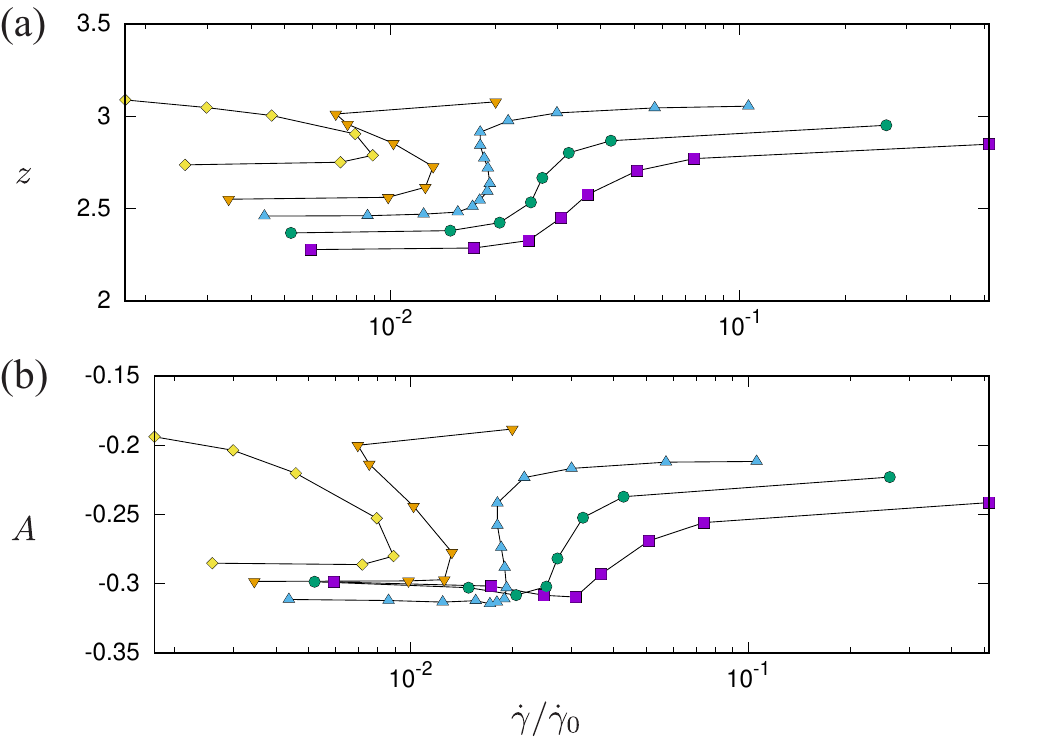}
\caption{
Semi-logarithmic plots of (a) $z$ and (b) $A$ as functions of the shear rate $\dot{\gamma}$,
where the symbols are as in Fig.\ \ref{fig:DST_et_df_xi_sz_sb-se}.
\label{fig:DST_sz_sb}}
\end{figure}

We also display $z$ and $A$ as functions of the shear rate $\dot{\gamma}$ (which is measured in simulations) in Figs.\ \ref{fig:DST_sz_sb}a and b, respectively.
%
\section{Dependence on the friction coefficient and system size}
\label{app:depend}
In this appendix, we examine the influence of the friction coefficient $\mu$ and system size (the number of particles) $N$
on the scaling relations,\ Eqs.\ (\ref{eq:delta_v}) and (\ref{eq:D_gd_eta}).

\begin{figure}
\includegraphics[width=\columnwidth]{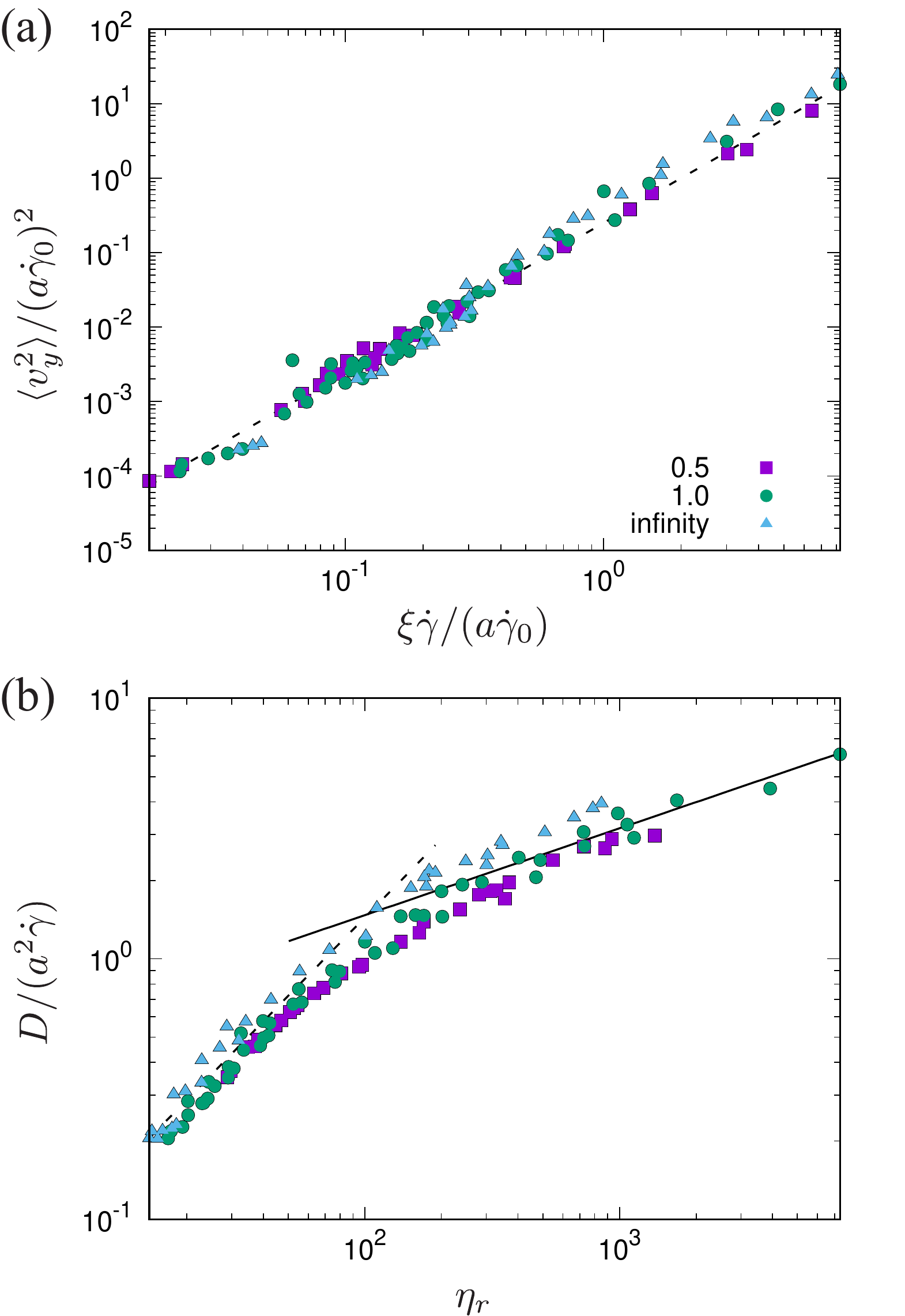}
\caption{
Scatter plots of (a) $\langle v_y^2\rangle$ and $\xi\dot{\gamma}$, and (b) $D/\dot{\gamma}$ and $\eta_r$,
where the system size is $N=2000$ and the friction coefficient $\mu$ increases as listed in the legend of (a) (``infinity" means $\mu=\infty$).
The lines in (a) and (b) are as in Figs.\ \ref{fig:DST_fluctuations_length} and \ref{fig:DST_viscosity_diffusivity}, respectively.
\label{fig:DST_appendix_mu-dependence}}
\end{figure}

Figure \ref{fig:DST_appendix_mu-dependence} displays scatter plots of (a) $\langle v_y^2\rangle$ and $\xi\dot{\gamma}$, and (b) $D/\dot{\gamma}$ and $\eta_r$,
where the system size is given by $N=2000$ and we vary $\mu$ as listed in the legend of (a).
Even though the friction coefficient affects the onset of DST, by controlling the $\phi_J$~\cite{Singh_2013, Singh_2018}.
As can be seen, the scaling relations (dashed and solid lines) well describe the data over the whole ranges of the packing fraction $\phi$ and shear rate $\dot{\gamma}$.
Therefore, Eqs.\ (\ref{eq:delta_v}) and (\ref{eq:D_gd_eta}) are not affected by the strength of friction coefficient $\mu$.

\begin{figure}
\includegraphics[width=\columnwidth]{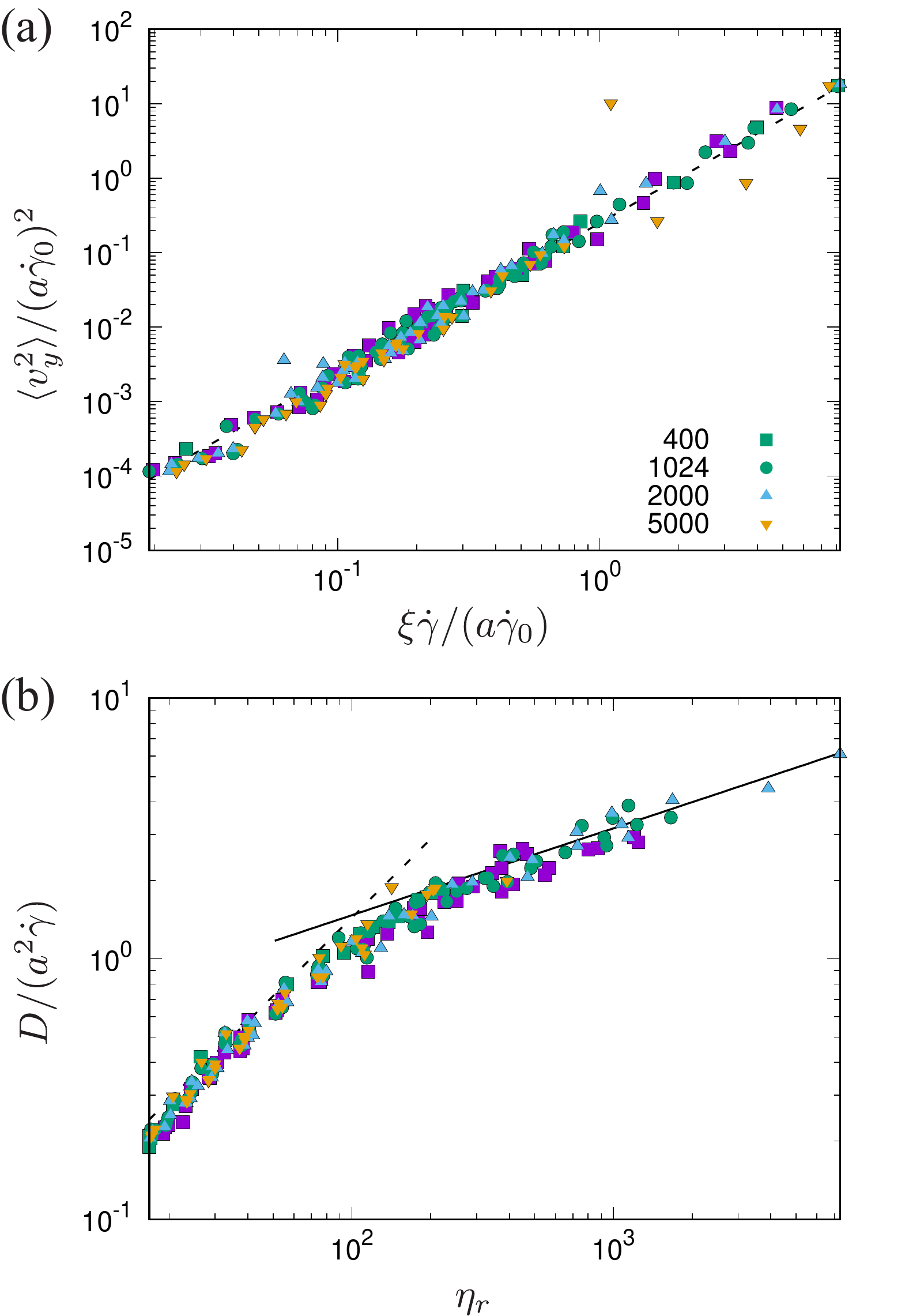}
\caption{
Scatter plots of (a) $\langle v_y^2\rangle$ and $\xi\dot{\gamma}$, and (b) $D/\dot{\gamma}$ and $\eta_r$,
where the friction coefficient is $\mu=1.0$ and the system size $N$ increases as listed in the legend of (a).
The lines in (a) and (b) are as in Figs.\ \ref{fig:DST_fluctuations_length} and \ref{fig:DST_viscosity_diffusivity}, respectively.
\label{fig:DST_appendix_N-dependence}}
\end{figure}

Figure \ref{fig:DST_appendix_N-dependence} shows the scatter plots of (a) $\langle v_y^2\rangle$ and $\xi\dot{\gamma}$, and (b) $D/\dot{\gamma}$ and $\eta_r$,
where $\mu=1$ and we increase $N$ as listed in the legend of (a).
As can be seen, all the data are well described by the scaling relations (dashed and solid lines)
so that Eqs.\ (\ref{eq:delta_v}) and (\ref{eq:D_gd_eta}) hold even if we change the system size $N$.
%
%
\end{document}